\documentclass[12pt]{iopart}
\usepackage{graphicx}

\begin{document}

\title[]{Study of $^{113}$In($\alpha,\alpha$) elastic scattering to determine $\alpha$-optical potential relevant for astrophysical $\gamma$-process}

\author{Dipali Basak$^{1,3}$, Tanmoy Bar$^{1,3}$, Lalit Kumar Sahoo$^{1,3}$, Sukhendu Saha$^{1,3}$, T. K. Rana$^{2,3}$, S. Manna$^{2,3}$, C. Bhattacharya$^{2,3}$, Samir Kundu$^{2,3}$, J. K. Sahoo$^2$, J. K. Meena$^2$, A. K. Saha$^2$, Ashok Kumar Mondal$^{4}$ and Chinmay Basu$^{1,3}$}

\address{$^1$ Saha Institute of Nuclear Physics, 1/AF, Bidhannagar, Kolkata$-$700064, India }
\address{$^2$Variable Energy Cyclotron Centre, 1/AF Bidhannagar,
Kolkata 700064, India}
\address{$^3$Homi Bhabha National Institute, Mumbai, Maharashtra$-$400094, India}
\address{$^4$Department of Physics, Manipal University Jaipur, Rajasthan$-$303007, India}

\ead{dipali.basak@saha.ac.in}
\vspace{10pt}
\begin{indented}
\item[]
\end{indented}

\begin{abstract}
 The $\alpha$-optical potential is one of the key input parameters used to measure the reaction rate of the ($\gamma,\alpha$)-process using the Hauser-Feshbach(HF) statistical model and the principle of detailed balance. $\alpha$-elastic scattering experiment on $^{113}$In $p$-nucleus was carried out in the energy range E$_{lab}$=26$-$32 MeV. The vacuum evaporation technique was used to prepare the $^{113}$In target~($\sim$86 $\mu$g/cm$^2$). An energy-dependent local optical potential parameters set was obtained by analysing the experimental elastic scattering angular distribution data. The local potential parameters are  extrapolated for lower energies and are used to measure the $^{113}$In($\alpha,\gamma$) reaction cross-section. 
\end{abstract}

%
\vspace{2pc}
%
%
%
%

\section{Introduction}
Above iron, the origin of the heavy and intermediate elements is not well understood. The bulk of these heavy nuclei are mainly synthesized by neutron capture and $\beta$-decay process. Elements are produced via either a slow($s$) neutron capture process or a rapid($r$) neutron capture process, based on the rate of neutron capture compared with $\beta$-decay~\cite{RevModPhys.83.157,argast2004neutron}. However, 30$-$35 naturally occurring neutron-deficient nuclei are not synthesized by the $s$- or $r$-process. The solar abundances of these nuclei are 10$-$100 times smaller compared to the $s$- or $r$-nuclei~\cite{woosley1978p,rayet1990p,arnould2003p}. The complete list of $p$-nuclei along with their solar abundances is given in Ref.~\cite{woosley1978p}. The production processes of these $p$-nuclei are not clearly comprehended. Different production sites for $p$-nuclei have been proposed in recent years~\cite{PhysRevC.73.015804}. Most of the $p$-nuclei are synthesized by a series of photodisintegration processes in explosive environments in type II and type Ia supernovae~\cite{travaglio2011type,travaglio2014radiogenic}.

In the photodisintegration process, $p$-nuclei are formed by ($\gamma$, n), ($\gamma$, p) or ($\gamma, \alpha$) process from the existing neutron-rich $s$- or $r$-seed nuclei, known as the $\gamma$-process. However, the photodisintegration process underproduces the $p$-nuclei in the Mo-Ru region and also in the 150$\leq$~A~$\leq$165 mass region~\cite{howard1991new,rauscher2002nucleosynthesis}. In order to explain the abundance of the $p$-nuclei, additional production mechanisms have been taken into consideration. Zs. Nemeth $et~al.$~\cite{nemeth1994nucleosynthesis} showed that the formation of some highly abundant $p$-nuclei ($^{113}$In, $^{115}$Sn, etc.) have a contribution from the $r$-process. Reaction network calculations are done for more than 10,000 nuclear reactions involving about 1000 stable and unstable nuclei to determine the abundance of $p$-nuclei. As unstable nuclei are experimentally inaccessible, reaction rates for stable nuclei are measured, and other reaction rates involving unstable nuclei are predicted using theoretical Hauser-Feshbach (HF) statistical model calculations~\cite{RAUSCHER20001,RAUSCHER200147}. HF calculation of cross-sections is sensitive to the choice of nuclear input parameters (optical model potentials, nuclear level densities, $\gamma$-ray strength functions). It has been observed from previous measurements that there are significant discrepancies between the theoretical predictions and measured value of the ($\gamma, \alpha$) reaction rates~\cite{rapp2001alpha, PhysRevC.74.025805, netterdon2013investigation, kiss2011determining}. The accurate knowledge of alpha optical potential at astrophysical relevant energy region can significantly minimize the discrepancy of the theoretical calculations. Determination of nuclear potential from scattering experiments at astrophysical energies is not possible, as Rutherford scattering is dominant at such energies, masking the nuclear effect. Therefore, elastic scattering experiments are performed at higher energies, and the potential parameters are extrapolated to the astrophysical energy region. The $^{113}$In($\alpha, \gamma$)$^{117}$Sb reaction is important to study as this is the inverse of $^{117}$Sb($\gamma, \alpha$)$^{113}$In reaction that contributes to the formation of $^{113}$In in the p-nuclei production site. The determination of $\alpha$-optical potential this nuclide is therefore crucial.

An elastic scattering experiment on $^{113}$In was carried out at E$_{lab}$~=~26, 29 and 32 MeV above the Coulomb barrier~(18 MeV). There is only one previous $^{113}$In($\alpha, \alpha$)$^{113}$In experiment at 16.1 and 19.5 MeV~\cite{PhysRevC.88.045804} that reported local potential parameters. The primary objective of this present work is to obtain an energy-dependent alpha optical potential in order to better understand the behaviour of particle fusion with A$\sim$100 nuclei.

\section{Experimental Procedure}
  The elastic scattering experiment was performed at K-130 Cyclotron VECC, Kolkata. Enriched $^{113}$In (93.7$\%$) targets on Al backing ($\approx$ 2.7 $\mu$m) were prepared by vacuum evaporation. The target thickness was obtained by measuring the energy loss of $\alpha$ particles from a known 3-line $\alpha$ source ($^{239}$Pu, $^{241}$Am, $^{244}$Cm). The measured thickness of the target was approximately $\approx$ 86 $\mu$g/cm$^2$, which is equivalent to 4.25$\times$10$^{17}$ atoms/cm$^2$. The uncertainty of the target thickness is about 6$-$7$\%$.

\begin{figure}
\begin{center}
\includegraphics[scale=1.0]{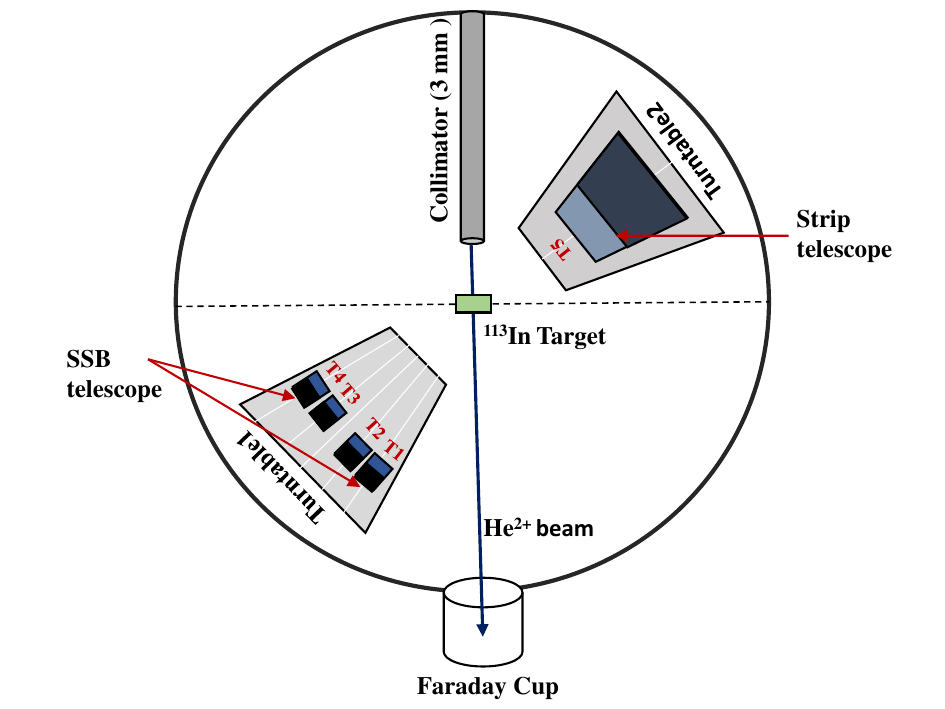}
\end{center}
\caption{Schematic diagram of the scattering chamber and detector setup during the experiment. A strip telescope (T5) was placed on a turntable to cover the backward angle (91$^{\circ}$-140$^{\circ}$), and four silicon surface barrier (SSB) telescopes (T1$-$T4) are placed on the another turntable to cover the forward angle (23$^{\circ}$-90$^{\circ}$).}\label{Fig1} 
\end{figure}

A scattering chamber of 90 cm diameter was used for the detector setup and measurements. The schematic diagram of the scattering chamber with the detector setup is shown in Fig~\ref{Fig1}. The targets were placed on a remote-controlled target ladder that was positioned in the centre of the scattering chamber. An alumina plate was also placed in one of the target frames on the ladder to monitor the beam diameter and position at the beginning and the end of each $\alpha$-beam energy. To determine the optical potential parameters, a well-defined beam energy is necessary. Thus, a collimator of 3 mm diameter was used throughout the measurement. Therefore, $\alpha$-beam energy resolution is about 200 keV, which contributes to the major uncertainty in the energy resolution of the spectra, as shown in Fig.~\ref{Fig3}. Elastic scattering cross-sections in angular range $\theta$ = 23$^{\circ}$-140$^{\circ}$  were measured at three different energies (E = 26, 29, 32 MeV). The current of the He$^{2+}$ beam during the experiment varied between $\sim$10-15 nA.

 Five $\Delta$E-E telescopes mounted on the two turntables inside the scattering chamber were used for particle identification and angular distribution measurements. Four silicon surface barrier (SSB) $\Delta$E-E telescopes (T1$-$T4) were placed in one turntable at relative angular separation of 10$^{\circ}$~(except for T2 and T3) to measure the cross-section at forward angles~(23$^{\circ}-$90$^{\circ}$). The relative angular separation between T2 and T3 telescopes was 20$^{\circ}$. SSB detectors had an active area $\approx$ 80 mm$^2$ and $\Delta$E-detector were 150~$\mu$m thick, and E-detectors had thicknesses between 500$-$3000~$\mu$m.
  
   A strip detector telescope (T5) was placed on the other turntable for backward angle measurements (91$^{\circ}-$140$^{\circ}$). The strip telescope consisted of a single-sided 16 channels silicon strip $\Delta$E-detector of thickness $\approx$ 52 $\mu$m and a double-sided 16 $\times$ 16 channels E detector having thickness $\approx$ 1034~$\mu$m. The relative angular separation of the strips was $\approx$ 1$^{\circ}$. The SSB telescopes were placed 262 mm, and the strip telescope was kept 186 mm from the target center. A circular slit of diameter $\sim$4 mm was used as a collimator for each SSB telescope. Each strip has a collimated area of 6$\times$3 mm$^2$. Solid angles for SSB telescope is 0.28 msr and 0.52 msr for strip telescope. Prior to the experiment, the 3-line alpha source was used to measure the energy resolution of the detector, and it was observed to be between 50 and 100 keV. The energy resolution was additionally impacted by the initial beam energy uncertainty from the Cyclotron and the energy loss inside the target material. The average uncertainty in the energy during the experiment was about $~$400 keV, as shown in the Fig.~\ref{Fig3}. The online data were collected using a VERSA Module Eurocard (VME) based online data acquisition system.
  
\section{Data Analysis and Results}
\begin{figure}[h]
\begin{center}
\includegraphics[scale=0.5]{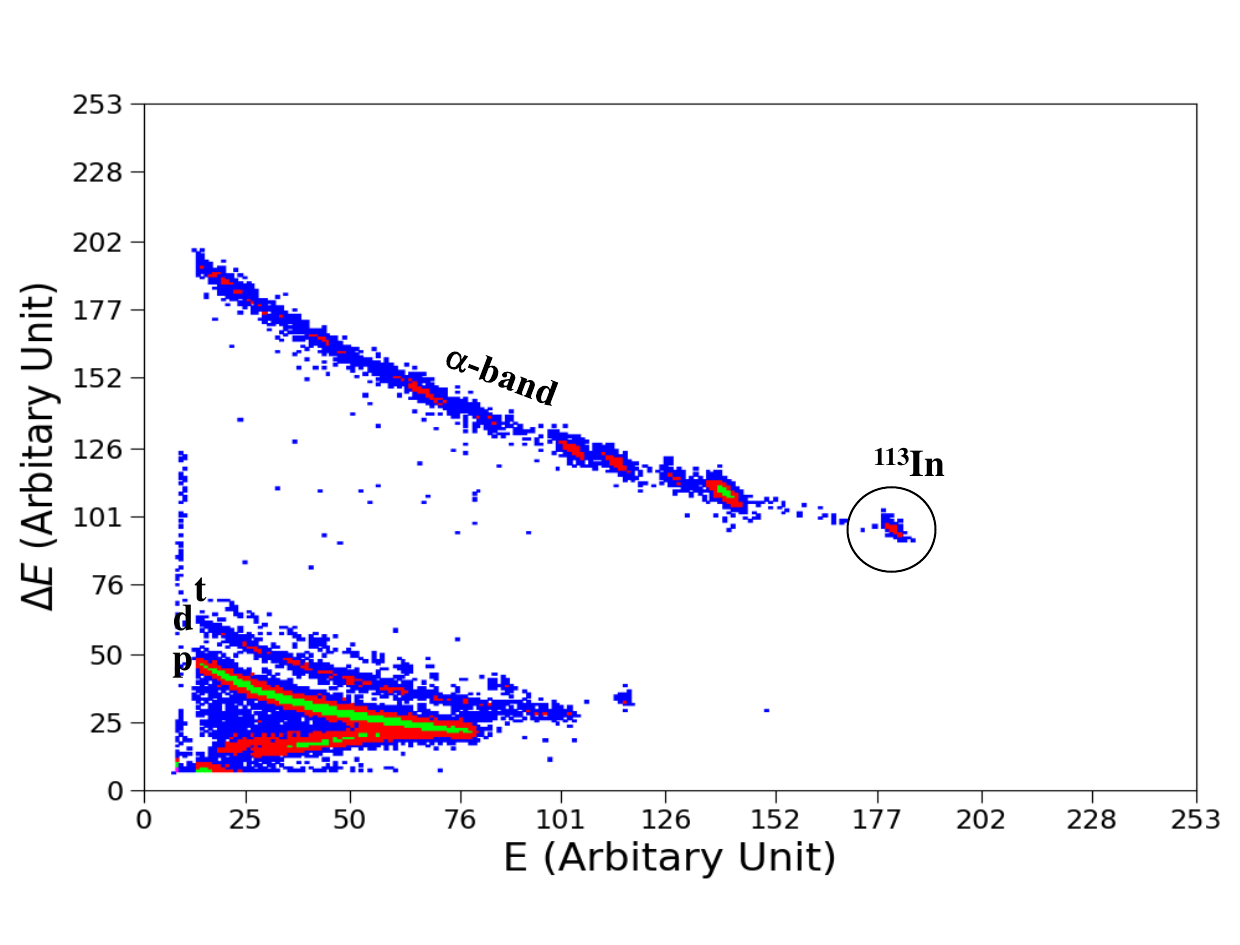}
\end{center}
\caption{2D $\Delta$E-E spectrum for $^{113}$In($\alpha, \alpha$)$^{113}$In elastic scattering at 29 MeV energy and at angle 66$^{\circ}$.}\label{Fig2}
\end{figure}

  The $^{113}$In($\alpha,\alpha$) elastic scattering angular distributions between $\theta$ = 23$^{\circ}-$140$^{\circ}$ were measured at three energies (E$_{lab}$ = 26, 29, 32 MeV) above the Coulomb barrier. The experimental 2D $\Delta$E-E spectrum  at 66$^{\circ}$ for 29 MeV $\alpha$-beam is shown in Fig.~\ref{Fig2}. The elastic counts corresponding to $^{113}$In($\alpha,\alpha$) elastic scattering are shown by a circle in the figure. Projection of the $\Delta$E-E spectrum on the x-axis yielded the 1D projection of the spectrum in terms of the ADC channel numbers. 
  
\begin{figure}[h]
\begin{center}
\begin{tabular}{cc}
\includegraphics[scale=0.3]{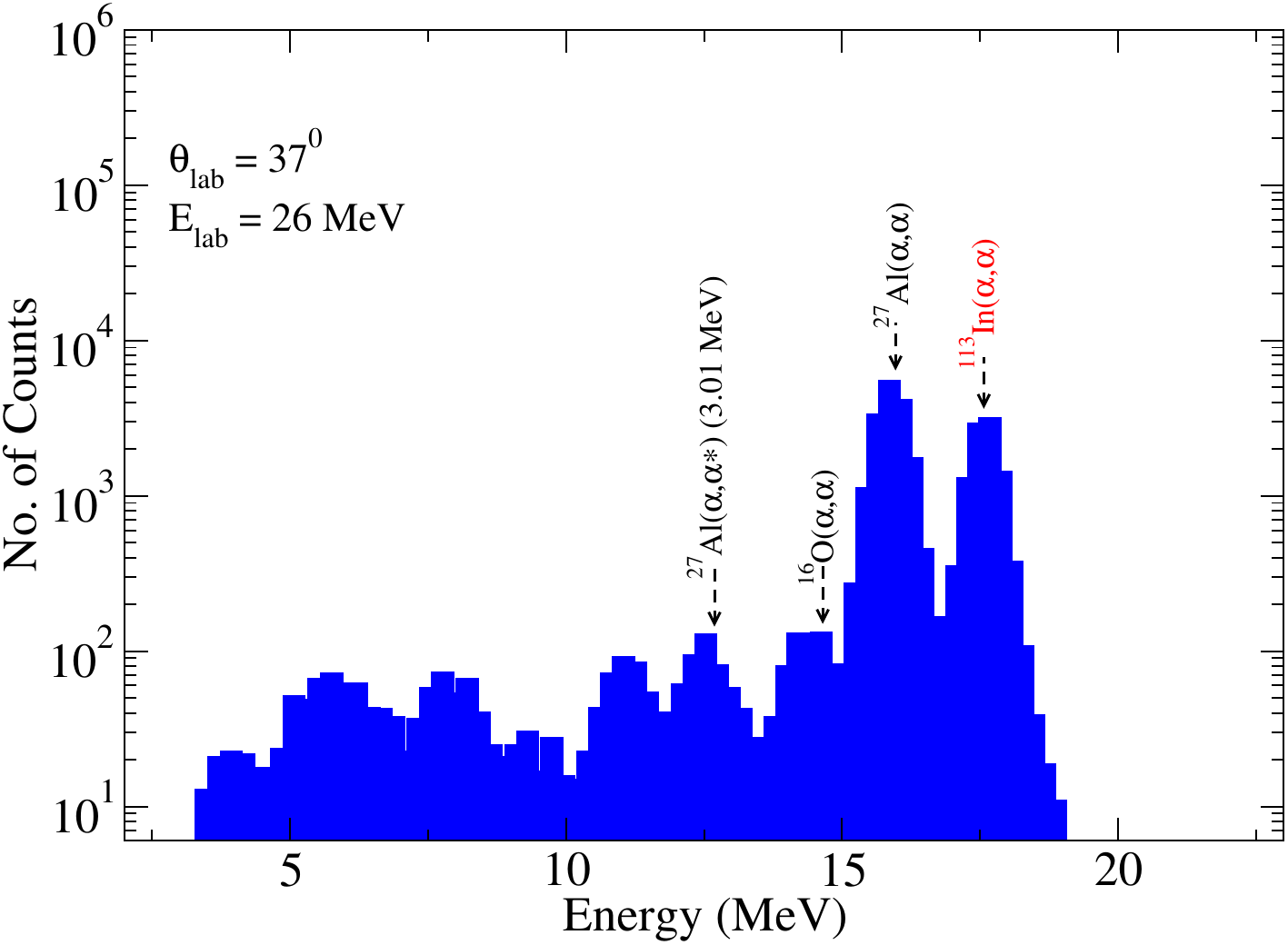}
\includegraphics[scale=0.3]{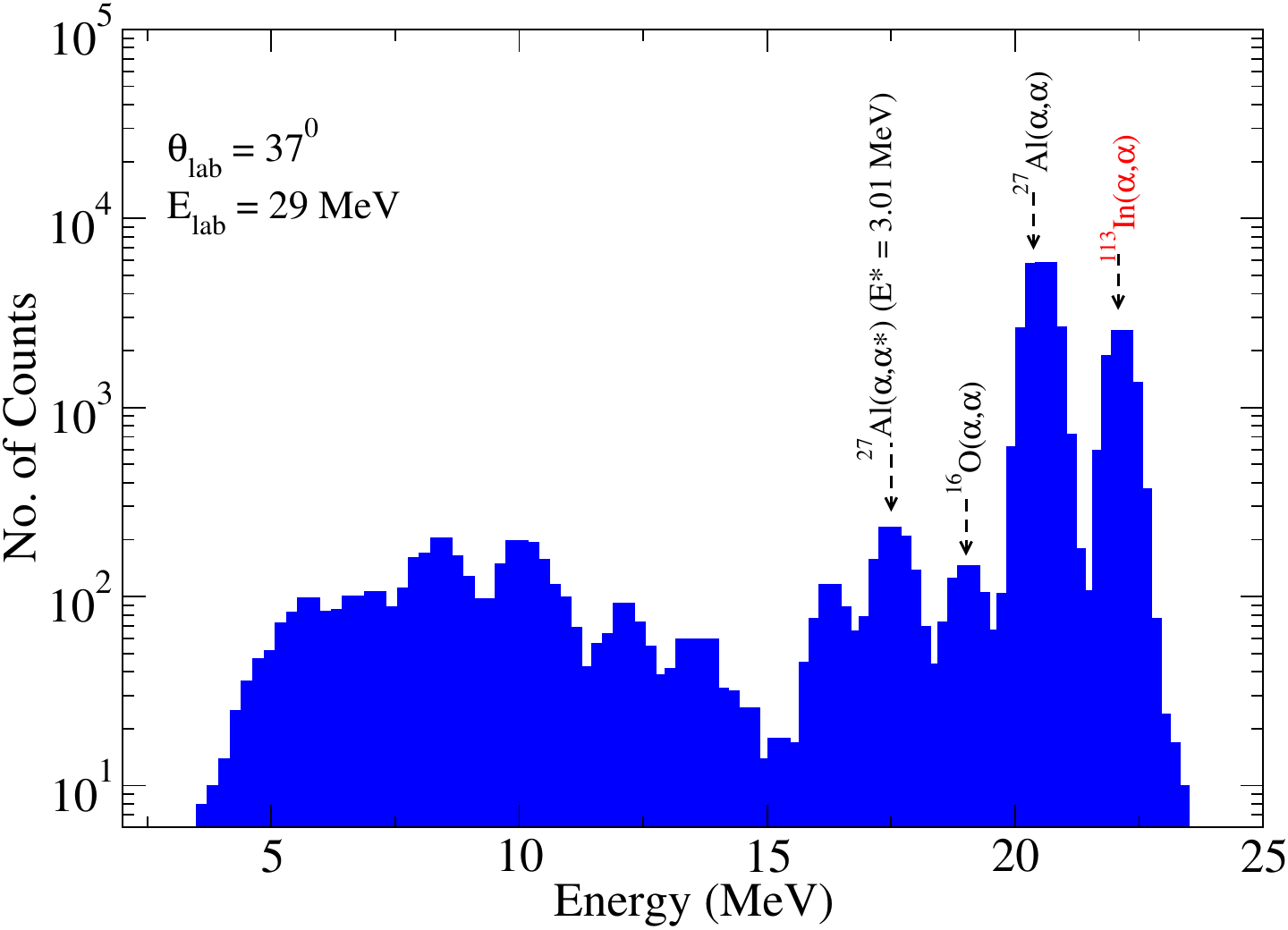}\\
\includegraphics[scale=0.3]{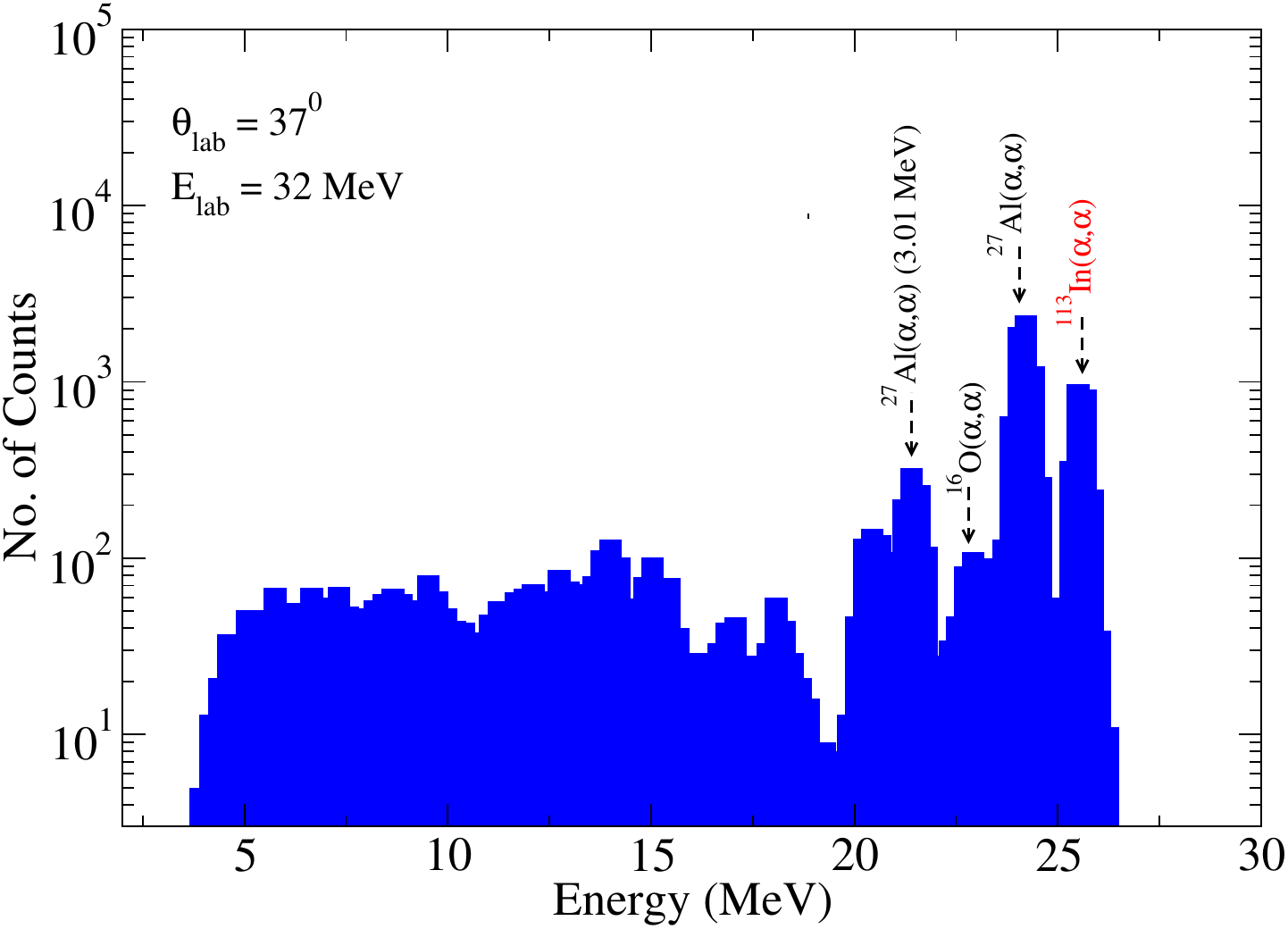}
\includegraphics[scale=0.3]{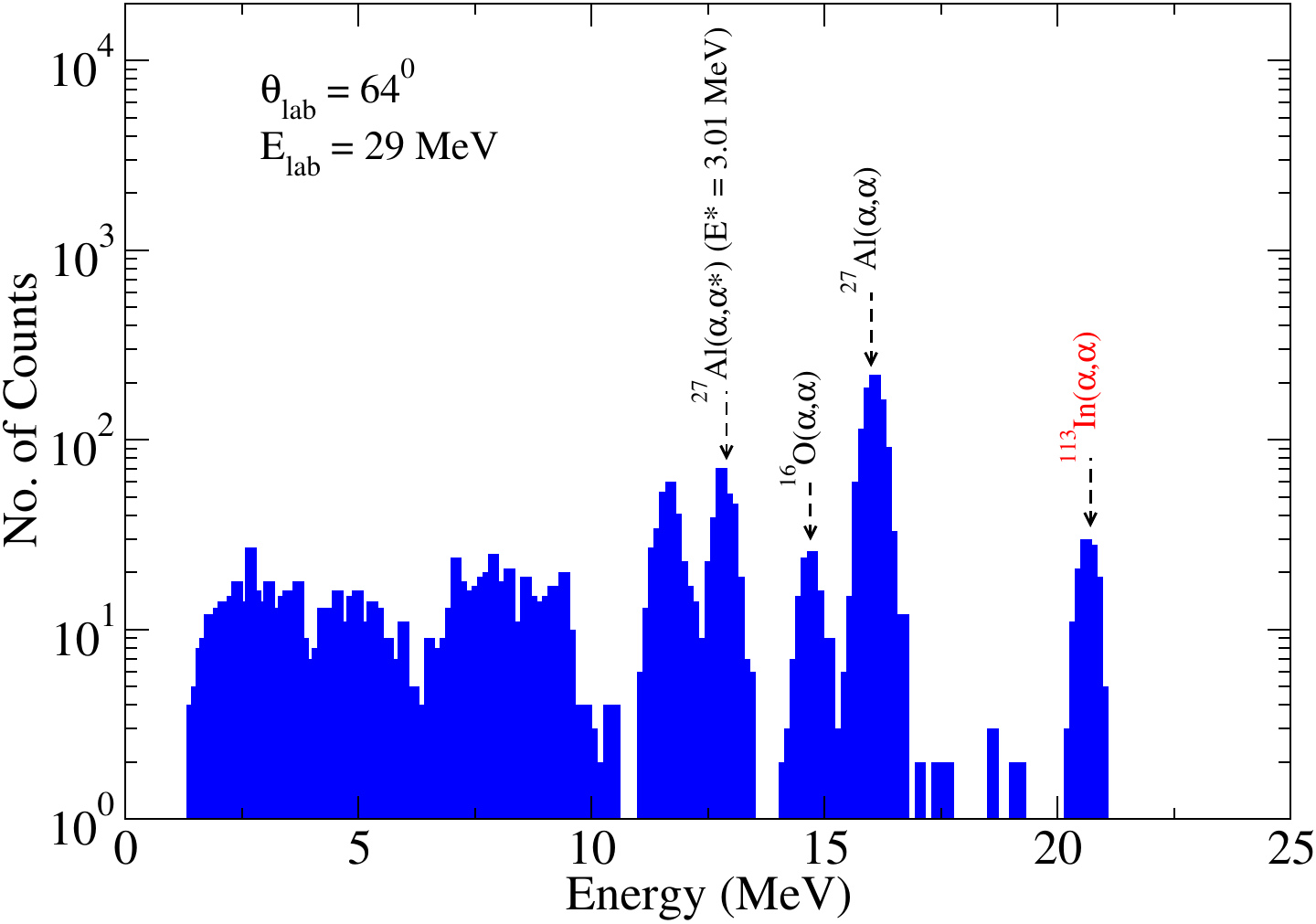}
\end{tabular}
\end{center}
\caption{Energy Spectra of $^{113}$In($\alpha, \alpha$) elastic scattering at $\theta_{lab}$ = 37$^{\circ}$ for three beam energies and at $\theta_{lab}$ = 64$^{\circ}$ for 29 MeV.}\label{Fig3}
\end{figure}

  Alpha elastic scattering with self-supporting $^{197}$Au target (230 $\mu$g/cm$^2$) at energies 26, 29, and 32 MeV were used to calibrate the energy spectrum. A $^{229}$Th source, with five different $\alpha$-energies viz. 4.9, 5.8, 6.3, 7.1, and 8.4 MeV was also used for low energy calibration. Energy spectra of scattered $\alpha$-particle at 37$^{\circ}$ and 64$^{\circ}$ angles for three beam energies are shown in Fig.~\ref{Fig3}. The differential cross-section for an angle is determined from the total count rate $A(\theta)$ of the scattered $\alpha$-particle from the target nucleus as,
  
  \begin{equation}
   \frac{d\sigma}{d\Omega} = \frac{A(\theta)}{N I\Omega}
  \end{equation}
where $I$ is the total number of $\alpha$-particles bombarding the target nucleus per unit time and $N$ is the surface density of the target material. $\Omega$ is the solid angle subtended by the detector at the target center. $A(\theta)$ is determined by calculating the required peak count of the  $\alpha$-spectrum using the CERN ROOT data analysis tool~\cite{brun1997root}.

Uncertainty in angular distribution data arises from systematic and statistical errors. The total count rate of scattered alpha particles from the target nucleus that have been detected in the telescope is the source of statistical uncertainty. For the forward angles, the average estimated statistical uncertainty was around 5$\%$. The count rate was much smaller at backward angles because of the forward peaked angular distributions and in case of measured angle coincided with the diffraction minima. The average backward angle statistical uncertainty  was around 15$\%$. 

Uncertainty in the measurement of the target thickness, the detector solid angle, and the calibration of the current digitizer were the primary sources of the systematic errors. The uncertainty of the target thickness was about 6$-$7$\%$ and for solid angle was about 4$\%$ and the uncertainty in the beam current was about 4$\%$. Due to the existence of  $^{115}$In in the target material, the uncertainty of the cross-section value of $^{113}$In($\alpha,\alpha$) was less than 1$\%$, as reported previously~\cite{PhysRevC.88.045804}. The inelastic contribution to the elastic $\alpha$-peak is an additional source of uncertainty. Spin-parity of first and second excited state of $^{113}$In are $\frac{1}{2}^-$ (0.39 MeV) and $\frac{3}{2}^-$ (0.65 MeV). The direct inelastic scattering to the first and second excited states of $^{113}$In was unlikely due to the parity change. The excitation of the third excited state $\left(\frac{5}{2}^+(1.02~{\rm MeV)}\right)$ was found to be very small ($\sim$0.38 mb at 32 MeV) by FRESCO calculation~\cite{thompson1988getting}. So, uncertainty due to the inelastic scattering was considered to be negligible in this analysis. The $\alpha$ peaks from backing Al and In were well separated above 45$^{\circ}$, and below 45$^{\circ}$, two peaks were separated by double Gaussian fitting using the CERN ROOT data analysis tool~\cite{brun1997root}. The total uncertainty was calculated as the quadratic sum of all systematic and statistical errors. 

\begin{figure}[h!]
\begin{center}
\begin{tabular}{cc}
\includegraphics[scale=0.3]{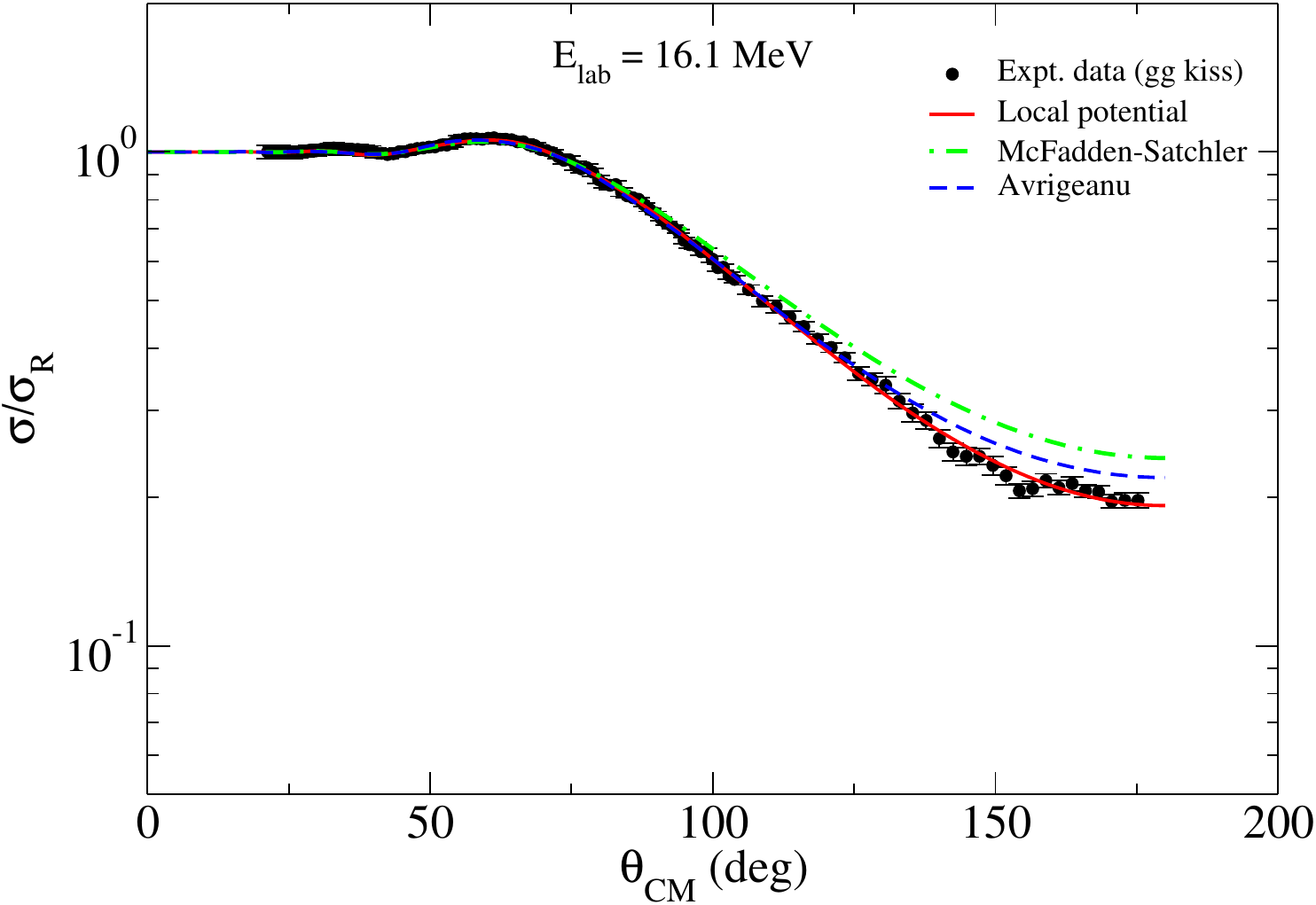}
\includegraphics[scale=0.3]{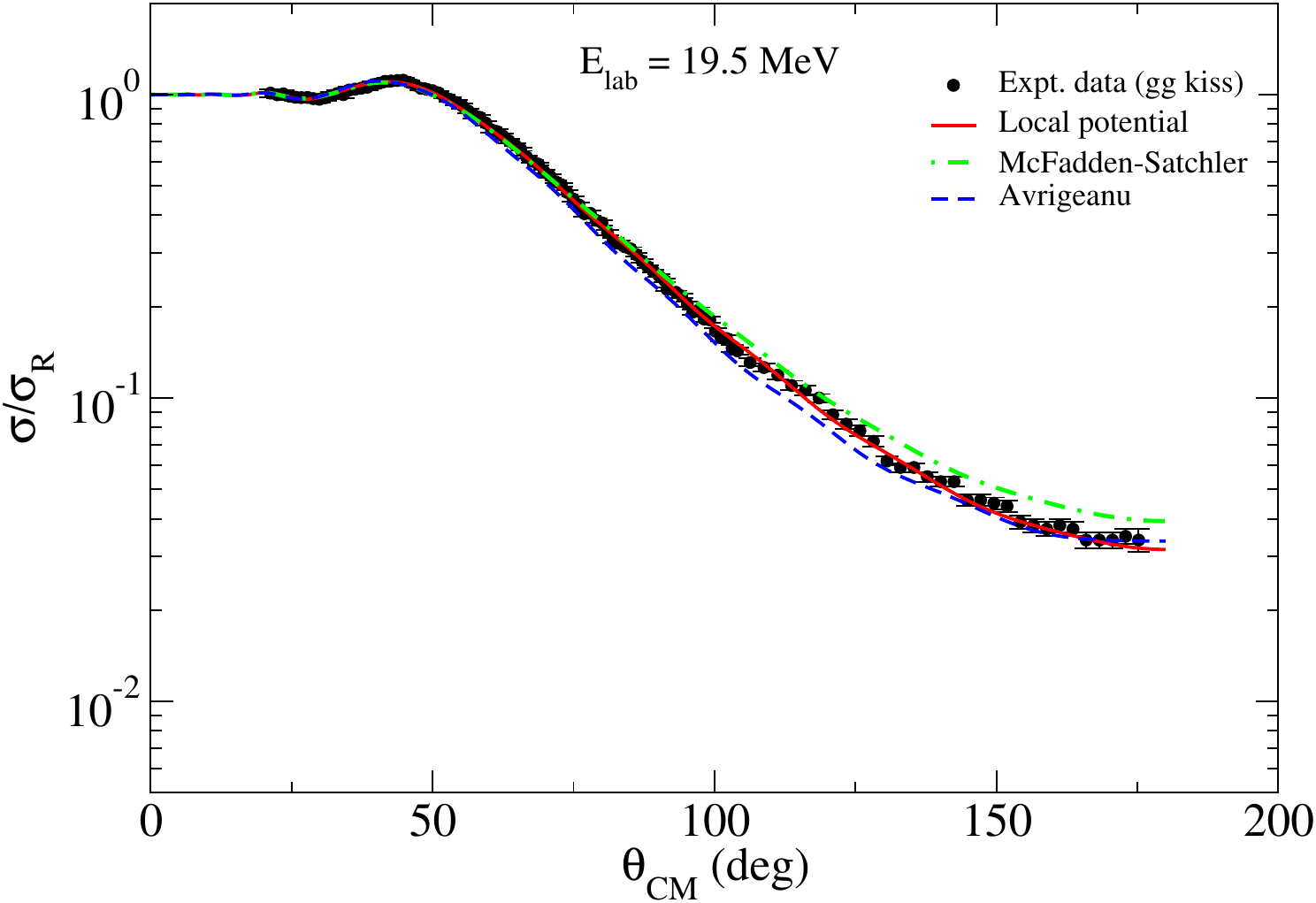}\\
\includegraphics[scale=0.3]{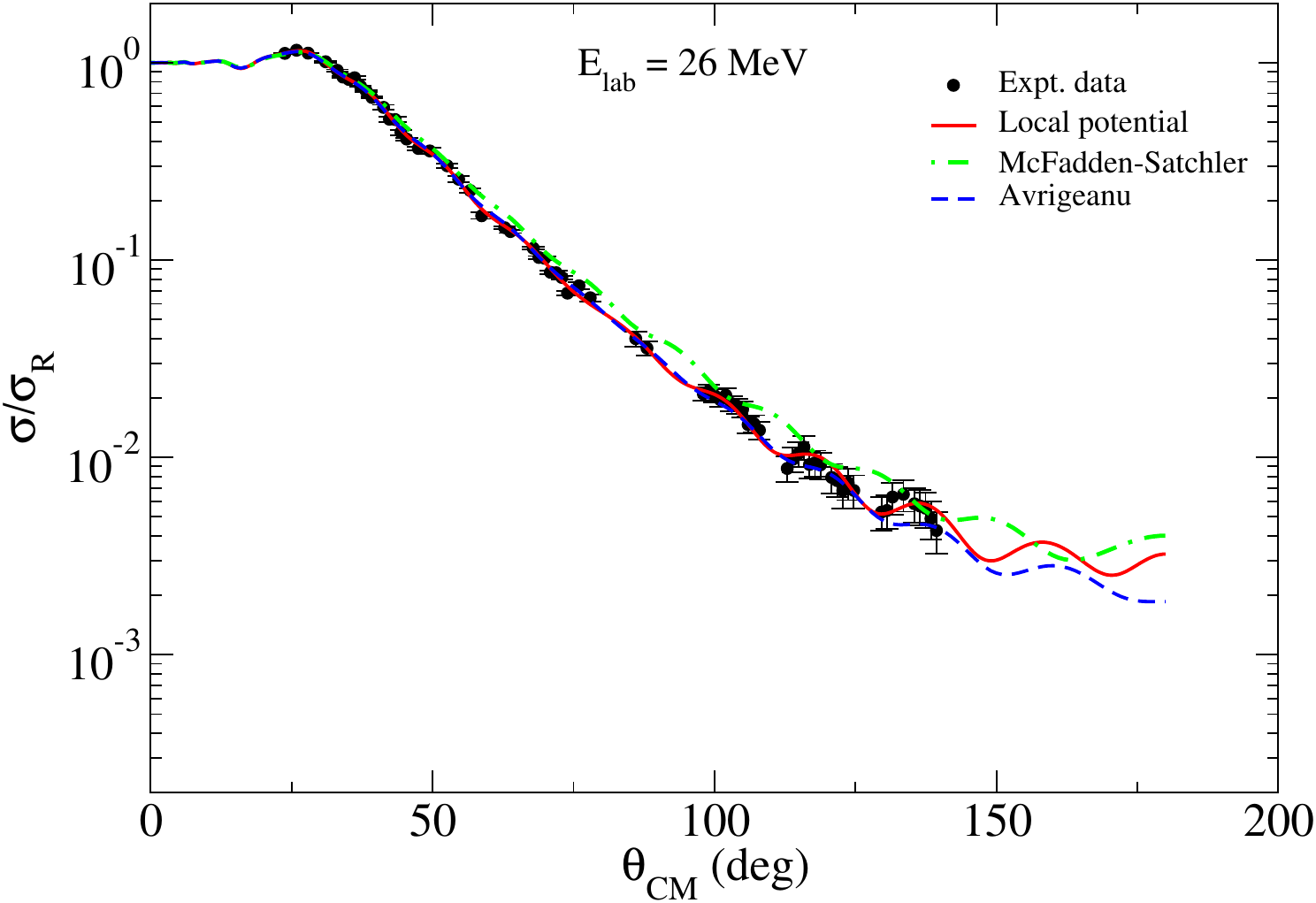}
\includegraphics[scale=0.3]{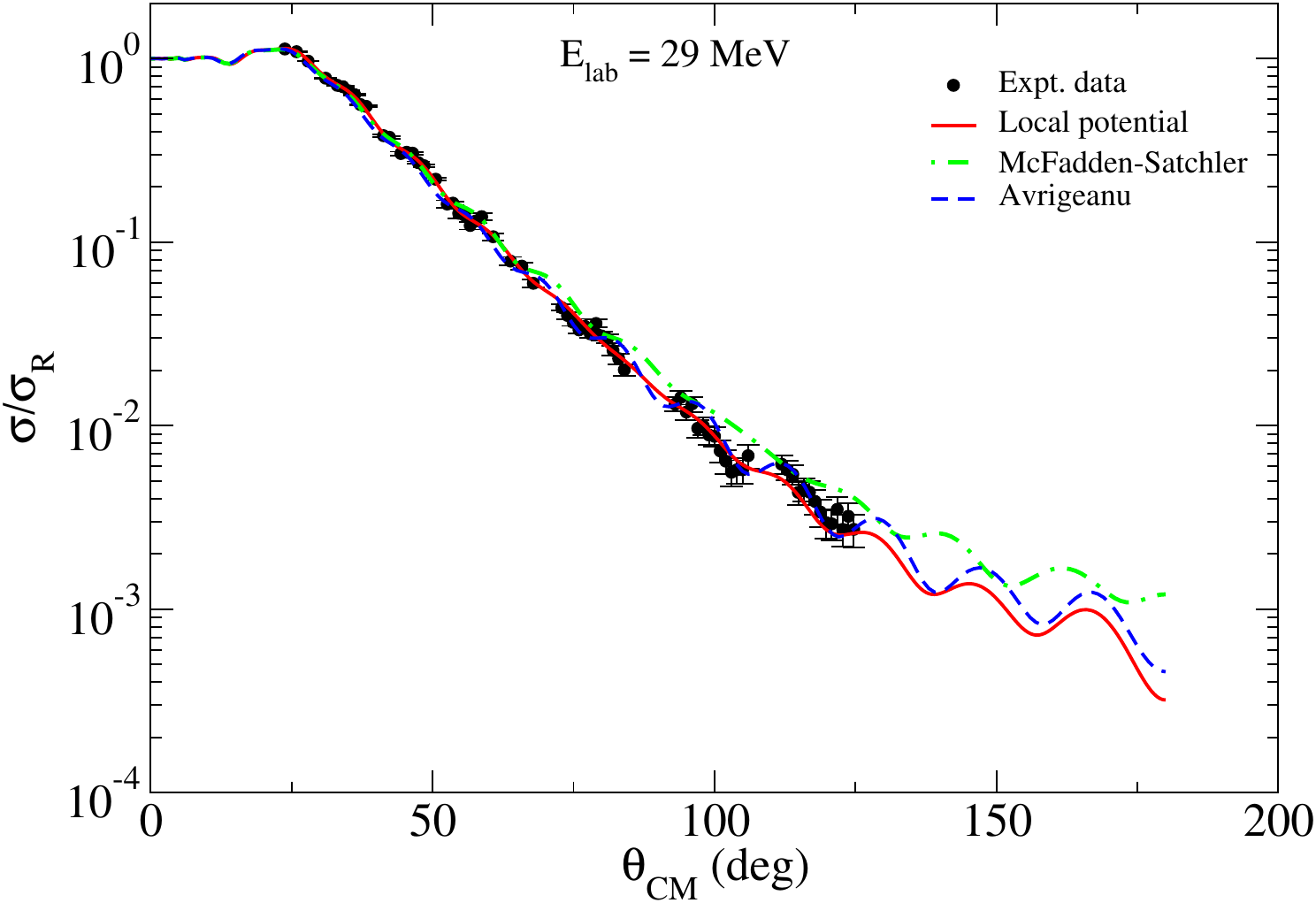}\\
\includegraphics[scale=0.3]{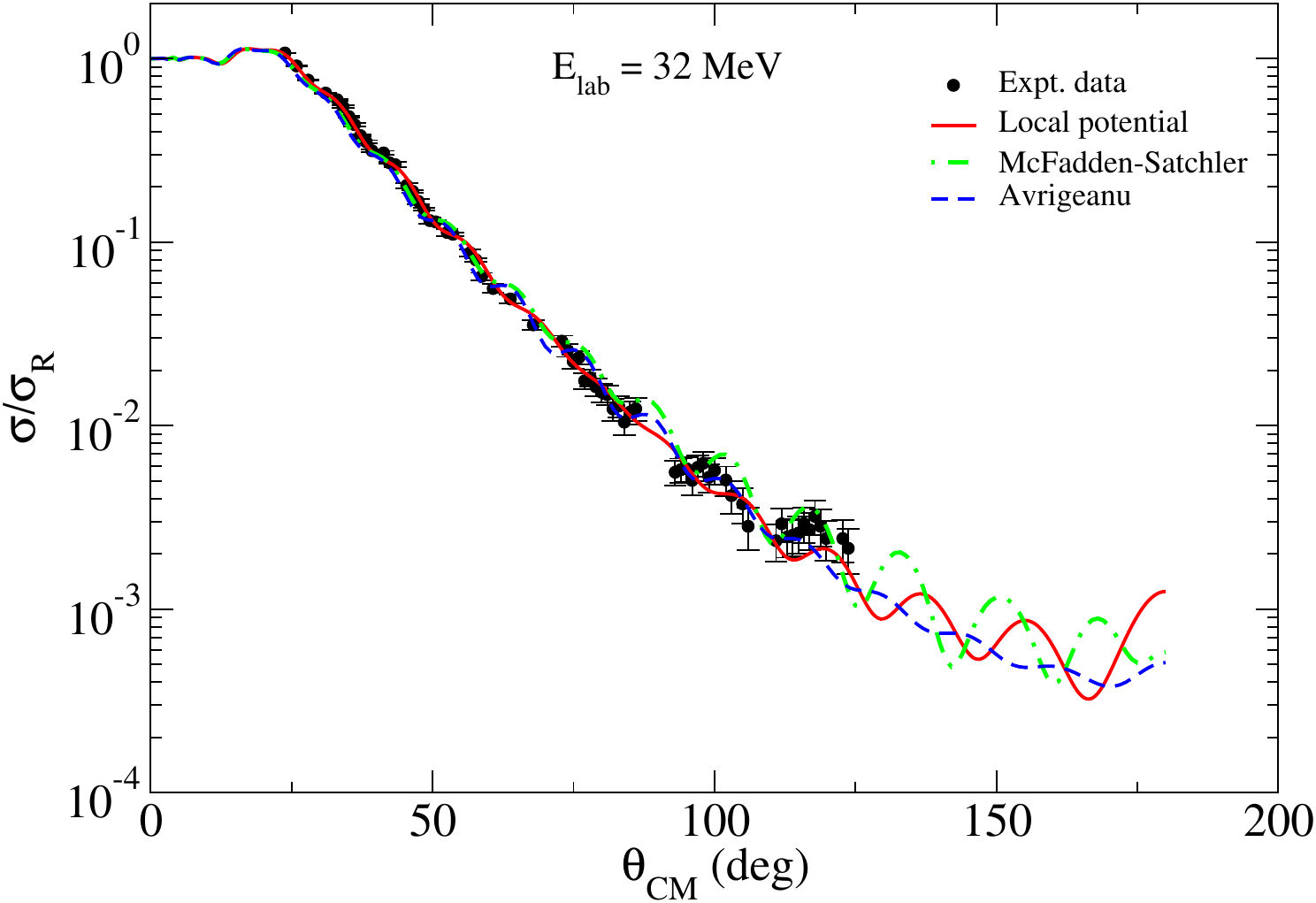}
\end{tabular}
\end{center}
\caption{Elastic scattering angular distribution at E$_{lab}$ = 16.1-32 MeV. Experimental data at E$_{lab}$ = 16.1, 19.5 MeV were taken from Ref.~\cite{PhysRevC.88.045804}. Theoretical calculations have been done with different global $\alpha$-optical potential.}\label{Fig4}
\end{figure}

The theoretical Rutherford cross-sections calculated from~\cite{tarasov2002code} were used to normalise the experimental differential cross-sections. The experimental angular distribution data from Ref.~\cite{PhysRevC.88.045804} have been included and performed the theoretical analysis in the energy range of 16.1 to 32 MeV. The experimental elastic scattering cross-sections are shown in Fig.~\ref{Fig4}.

\subsection{Local potential paramerization}
In the framework of the optical model~\cite{feshbach1958optical} a complex potential in combination of Coulomb ($V_{\rm C}$) and nuclear potential ($V_{\rm N})$ is given by

\begin{equation}
U(r)=V_{\rm C}(r) + V_{\rm N}(r).
\end{equation}

The Coulomb potential $V_{\rm C}(r)$ is calculated from the uniformly charged sphere of radius $R_{\rm C}$ ( $R_{\rm C}$ = $r_{\rm C}$($A_{\rm t}^{1/3}$ + $A_{\rm p}^{1/3}$)).
The nuclear potential consists of a real ($V(r)$) and a imaginary part ($W(r)$). Real part of nuclear potential is parameterized by volume Wood-Saxon potential and for imaginary part both volume and surface Wood-Saxon potential parametrization have been considered,

\begin{equation}
\eqalign{V(r)=-V_{\rm 0}f_{\rm v}(r), \cr
W(r)=-\left( W_{\rm 0}f_{\rm w}(r)+4a_{\rm s}W_{\rm s}\frac{ df_{\rm s}}{ dr}\right)} 
\end{equation}
where $V_{\rm 0}$ ,$W_{\rm 0}$ , $W_{\rm s}$ are the potential depths of volume real, volume imaginary and surface imaginary respectively. Wood-Saxon form factor is given by 
 \begin{equation}
  f_{\rm i}(r)=\left({1+{\rm exp}\frac{r-R_{\rm i}}{a_{\rm i}}}\right)^{-1}
 \end{equation}
where, \begin{eqnarray}
R_{\rm i} = r_{\rm i} (A_{\rm t}^{1/3} + A_{\rm p}^{1/3})
\qquad& \rm i = \rm v, \rm w, \rm s 
\end{eqnarray}

$r_{\rm i}$ and $a_{\rm i}$ are the radii and diffusivities of the each potential.

A new set of potential parameters was obtained by fitting the angular distribution of differential cross-section data using the search code SFRESCO~\cite{thompson1988getting}. The local potential parameters with $\chi^2$ per degree of freedom ($\chi^2_{red}$) values are listed in Table~\ref{tbl1} and calculations are shown by red solid line in Fig.~\ref{Fig4}.

\begin{table}[ht]
  \begin{center}
  \caption{Local optical potential parameters set obtained from experimental elastic scattering angular distribution fitting }\label{tbl1}
   \begin{tabular}{p{0.08\textwidth}cccccccccp{0.12\textwidth}} 
    \br
        \textbf{E$_{lab}$} & \multicolumn{3}{c}{\textbf{Volume real}} &\multicolumn{3}{c}{ \textbf{Volume imaginary}} & \multicolumn{3}{c}{ \textbf{Surface imaginary}}
      & \textbf{$\chi^2_{red}$} \\
     (MeV) & $V_{\rm 0}$ & $r_{\rm v}$ & $a_{\rm v}$ & $W_{\rm 0}$ & $r_{\rm w}$ & $a_{\rm w}$ & $W_{\rm s}$ & $r_{\rm s}$ & $a_{\rm s}$ & \\
      
      \mr
            16.1$^a$ & 63.04 & 1.17 & 0.50 & 8.00 & 1.16 & 0.42 & 14.01 & 1.16 & 0.26 & 0.49 \\ 
     19.5$^a$ & 57.7 & 1.16 & 0.49 & 8.61 & 1.16 & 0.42 & 12.84 & 1.11 & 0.25 & 1.03 \\
     26 & 62.12 & 1.15 & 0.53 & 9.33 & 1.16 & 0.42 & 8.69 & 1.08 & 0.29 & 1.8 \\
     29 & 48.16 & 1.14 & 0.54 & 9.57 & 1.16 & 0.42 & 5.8 & 1.06 & 0.36 & 2.4 \\
     32 & 54.24 & 1.11 & 0.56 & 10.04 & 1.16 & 0.42 & 5.53 & 1.09 & 0.36 & 2.2 \\ 
     \br 
   
             \end{tabular}
             $^a$ Experimental angular distribution data taken from Ref.~\cite{PhysRevC.88.045804}
             \end{center}
             \end{table}
             
  In this study, an energy-dependent parametrization of the local alpha potential is proposed~(Table~\ref{tbl2}) based on the best fit of the elastic scattering angular distribution data in the energy range of 16.1$-$32 MeV.  

\begin{table}
  \begin{center}
  \caption{Energy dependent local optical potential parameters}\label{tbl2}
  \begin{tabular}{cc}
  \br
  \textbf{Potential} & \textbf{Potential parameter}\\
  \mr
Volume real & $V_{\rm 0}=69.54-0.385{\rm E}-0.0048{\rm E}^2$ \\
& $r_{\rm v}=1.03+0.014{\rm E}-0.00037{\rm E}^2$ \\
& $a_{\rm v} = 0.60-0.011{\rm E}+0.00031{\rm E}^2$ \\
\mr
  Volume imaginary & $W_{\rm 0}=5.43+0.184{\rm E}-0.0013{\rm E}^2$ \\
  &$r_{\rm w}=1.16$ \\
  & $a_{\rm w}= 0.42$ \\
  \mr
 Surface imaginary & $W_{\rm s} =24.43-0.646{\rm E}+0.0013{\rm E}^2$ \\
 & $r_{\rm s}=1.62-0.041{\rm E}+0.00075{\rm E}^2$ \\
 & $a_{\rm s}=0.26-0.004{\rm E}+0.00023{\rm E}^2$ \\
  \mr
  Coulomb & $r_{\rm C}~=~1.25$\\
  \br 
  \end{tabular}
  \end{center}
  Energy (E) measured at laboratory frame. $V_{\rm 0}$, $W_{\rm 0}$, $W_{\rm s}$ are in MeV and $r_{\rm C}$, $r_{\rm i}$, $a_{\rm i}$ are in fm.
  \end{table}    

\subsection{Global optical potential}
Several global $\alpha$-optical model potentials have been proposed in recent years.  McFadden-Satchler~\cite{mcfadden1966optical} and Avrigeanu~\cite{avrigeanu2014further} are the two most widely used
global $\alpha$-optical model potentials. Both the real and imaginary nuclear potential in these two potential models can be expressed in Wood-Saxon potential form. Elastic scattering cross-sections were calculated using the global optical potential with the code FRESCO and compared with the experimental experimental data. $\chi^2_{red}$ were calculated for these two potentials and values were listed in Table~\ref{tbl3}.
\begin{table}[h!]
  \begin{center}
  \caption{ $\chi^2_{red}$ values of $^{113}$In($\alpha,\alpha$) scattering using global optical potentials}\label{tbl3}
   \begin{tabular}{ccc} 
    \br
        \textbf{E$_{lab}$} (MeV) &  \textbf{McFadden-Satchler} 
      & \textbf{Avrigeanu} \\
      \mr
       16.1 & 9.85 & 2.53 \\
       19.5 & 7.32 & 17.54 \\
      26 & 17.83 & 2.49 \\
      29 & 12.56 & 20.12 \\
      32 & 17.87 & 29.28 \\
     \br   
   
             \end{tabular}
             \end{center}
             \end{table}
             
\subsection{Alpha induced reaction on $^{113}${\rm In}}
\begin{figure}
\begin{center}
\includegraphics[scale=0.31]{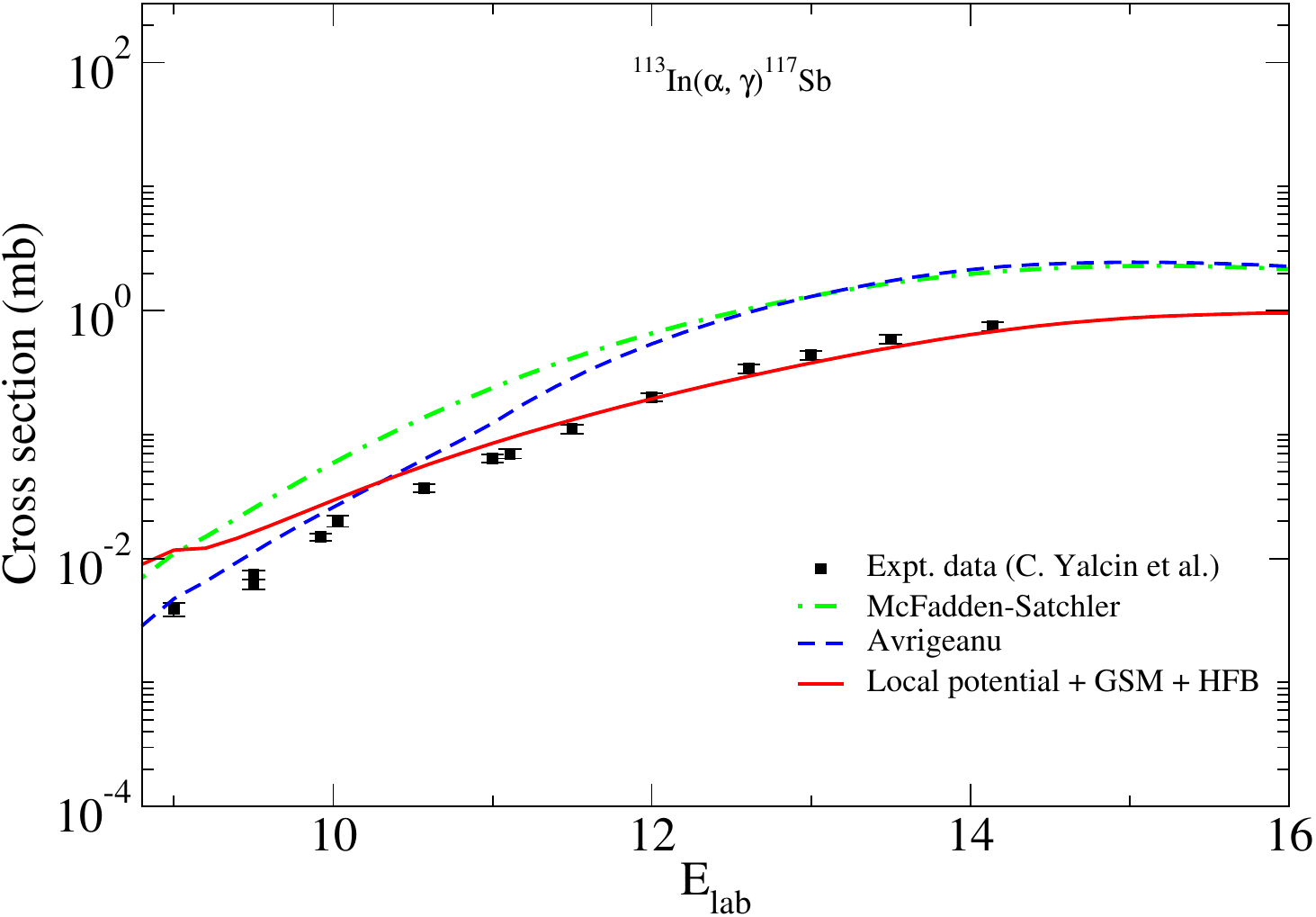}
\end{center}
\caption{($\alpha, \gamma$) reaction cross-section. Experimental data from Ref.~\cite{PhysRevC.79.065801} compared to the theoretical predictions using local optical potential.}\label{Fig5}
\end{figure}

 The derived energy-dependent local alpha potential parameters listed in Table~\ref{tbl2} are extrapolated for lower energies and used to calculate the $^{113}$In($\alpha,\gamma$) reaction cross-section at the astrophysical energy region. The theoretical prediction obtained from the new local potential is compared with the previously measured ($\alpha,\gamma$) cross-sections~\cite{PhysRevC.79.065801}, as shown in Fig~\ref{Fig5}. The widely used global $\alpha$-optical potentials McFadden-Satchler and Avrigeanu were also used in theoretical calculations in order to compare the results with the local potential. Statistical model code TALYS-1.96~\cite{koning2023talys} was used for theoretical analysis. It was observed that the global potentials overestimate the reaction cross-sections. However, the new local potential better explains reaction data with generalised superfluid model~\cite{ignatyuk1979role,ignatyuk1993density} for level density and Hartree-Fock-Bogoliubov~\cite{goriely2004microscopic} for $\gamma$-ray strength function. The fitting with local potential was better at higher energy region but deviated at lower energies.

\section{Discussions}
Elastic scattering angular distributions for $^{113}$In $p$-nuclei were measured at three energies above the Coulomb barrier (26, 29 and 32 MeV). The elastic scattering angular distribution data from the present and previous measurements~\cite{PhysRevC.88.045804} were fitted. An energy-dependent local potential was obtained for $^{113}$In nucleus from best fitting. The two global $\alpha$-optical potentials, McFadden-Satchler~\cite{mcfadden1966optical} and Avrigeanu~\cite{avrigeanu2014further} in the studied energy range cannot adequately describe the experimental elastic scattering angular distribution data, and the $\chi^2_{red}$ values are relatively poor than the local potential fitting. The obtained local $\alpha$-optical model potential satisfactorily explains the ($\alpha, \gamma$) reaction cross-section data than the global optical model. However, the local optical potential overestimates the ($\alpha, \gamma$) reaction data at lower energy region. ($\alpha, \gamma$) reaction cross-sections close to the Gamow window~(6.13$-$9.03 MeV at T$_9$=3~\cite{PhysRevC.81.045807}) are astrophysically more important. Such theoretical overestimation of the ($\alpha, \gamma$) cross-section at lower energies indicates the presence of additional factors that suppress cross-section value. Similar behaviour at lower energies was also observed for other p-nuclei~\cite{PhysRevC.85.035808, ornelas2015106cd}. As ($\alpha, \gamma$) reaction also depends on other nuclear input parameters such as level density, $\gamma$-ray strength function, etc., accurate estimation of these parameters is needed to solve the low energy anomaly.

The $^{113}$In is considered in a spherical shape to simplify the calculations. Further studies considering the deformed shape of the nuclei might be useful for better description. A new parameterization
for the $\alpha$-optical potential may be beneficial to accurately describe the low energy reaction data.  Recently, ($\alpha$, n) reaction cross-sections have been used to obtain the $\alpha$-optical potential suitable for low energy reactions~\cite{basak2022determination, PhysRevC.105.014602}. More experimental measurements of elastic scattering and ($\alpha$, n) reaction are needed to obtain better potential parametrizations that can describe reaction data relaibly in astrophysical energy regime.

\section*{Acknowledgement}
Author would like to thank the VECC Cyclotron facility for their help during the experiment. Author also acknowledge FRENA target facility for preparation of enriched $^{113}$In target and Mr. Sudipta Barman and other workshop members of Saha Institute of Nuclear Physics, Kolkata for their kind support. SS would acknowledge the Council of Scientific and Industrial Research (CSIR), Government of India, for funding assistance (File No 09/489(0119)/2019-EMR-I).
\section*{References}

\bibliography{Dipali_jpg}

\end{document}